\documentclass[aps,twocolumn,nofootinbib]{revtex4}

\usepackage{graphicx}
\usepackage{amssymb, amsmath,mathtools}
\usepackage{bbm}
\usepackage{enumerate}
\usepackage{xcolor}

\graphicspath{{./}{./plots/}}

\usepackage{xcolor}

\begin{document}

\title{Symplectic symmetry of quadratic-band-touching Hamiltonians in two dimensions}

\author{Igor F. Herbut and Samson C. H. Ling}

\affiliation{Department of Physics, Simon Fraser University, Burnaby, British Columbia, Canada V5A 1S6}

\begin{abstract} The internal low-energy symmetry of the massless Lorentz-invariant Dirac Hamiltonian in $2+1$ dimensions is known to be $O(2N)$, where $N$ is the number of two-component Dirac fermions. Here we point out that there exists an analogous internal symmetry of the single-particle quadratic-band-touching Hamiltonian in two spatial dimensions, and it is the unitary symplectic group, $USp(2N)$. All fermion bilinears belong to one of the three small irreducible representations of this group. The interacting theory that respects the $USp(2N)$ symmetry and the spatial rotations is constructed and found to allow two independent interaction terms. When these interactions are infrared-relevant the symplectic symmetry either remains preserved or becomes spontaneously broken to $USp(N) \times USp(N)$. The symmetry in lattices such as honeycomb to infinite order in the dispersion's expansion in powers of local momentum is given by the overlap of the symplectic and the orthogonal groups. We show that this overlap is $O(2N) \bigcap USp(2N) = U(N)$. 
Some consequences of the symplectic symmetry and its spontaneous breaking are briefly discussed.      
\end{abstract}

\maketitle

\section{Introduction}

Graphene, or any other two-dimensional Dirac material, famously features two inequivalent non-trivial Fermi points (or valleys) in the momentum space near which the gapless excitations acquire quasi-ultra-relativistic energy-momentum dispersion. This is a well-appreciated consequence of the time-reversal symmetry and the Nielsen-Ninomiya theorem. \cite{Nielsen}  As a consequence, at low energies and weak interactions a new $SU(2)$ symmetry of ``valley rotations" emerges, which together with the usual $SU(2)$ symmetry of spin rotations and the particle-number $U(1)$ becomes a part of the emergent internal (or ``flavor") $U(4)$. In real graphene this larger symmetry is weakly broken by the electron-electron interaction and the higher-order corrections to the energy-momentum dispersion that derive from the underlying lattice. At strong interactions or in finite magnetic field one of many competing ordered states may emerge as the ground state. \cite{HerbutReview}  

It is a mathematical fact, however, that the $U(N)$-symmetric massless Dirac Hamiltonian in $2+1$ space-time dimensions, with electron-electron interactions and dispersion's non-linearities neglected, is secretly $O(2N)$-symmetric. \cite{HerbutMandal} This is also true in $1+1$ dimensions, \cite{ZinnJustin} and it is maybe most easily seen by writing the usual  Grassmann field for the complex fermion in terms of its real and imaginary (``Majorana") components, and then by choosing the basis in which the two Pauli matrices in the Dirac Hamiltonian are both real. In this basis the two Majorana components decouple, and the $O(2N)$  symmetry, larger than the original $U(N)$, readily emerges. One can then proceed to identify all fermion bilinears as members of irreducible representations (``irreps") of the $O(2N)$. The Lorentz-invariant mass-terms, in particular, fall either into the singlet or the symmetric two-component tensor representation.\cite{HerbutMandal}  The  interaction term that would respect $O(2N)$ turns out to be unique. Consequently, any Lorentz-invariant interacting field theory in $2+1$ dimensions with the quartic interaction term that respects $O(2N)$ symmetry is equivalent to the Gross-Neveu model.\cite{HerbutMandal, GrossNeveu, Han}  

In going from Dirac to Majorana representation the emergence of the larger orthogonal group seems almost inevitable. In this note we show, however, that this is not so, and that its appearance is dependent on the linearity, or more precisely, on the oddness of the Dirac Hamiltonian under momentum reversal. Assuming the single-particle Hamiltonian of $N$ two-component complex fermions in two spatial dimensions to be even under momentum reversal one finds that the symmetry becomes $USp(2N)$, the unitary symplectic group. \cite{Zee} The fermion mass-like bilinears still fall into two distinct irreps of $USp(2N)$ of dimensions $1$ and $ N (2N-1)-1$. \cite{irrep} In contrast to the Dirac case, however, there now exist {\it two} Fierz-independent interaction terms that respect the $USp(2N)$ and the $O(2)$ of the spatial rotations. When the corresponding couplings are infrared-relevant in the renormalization group sense, the $USp(2N)$ ends up being either preserved or spontaneously broken to $USp(N) \times USp(N)$ in the ground state.   

When expanding around the Fermi points of the honeycomb lattice in powers of local momentum one naturally finds both the terms that are even and the terms that are odd under momentum inversion. Since all the odd terms share the same $O(2N)$ symmetry whereas the even terms share the $USp(2N)$, the overall symmetry to all orders in such an expansion is given by the overlap between these two groups. We show that this overlap is yet another $U(N)$.    

The paper is organized as follows. In the next section we discuss the emergence of the orthogonal symmetry in more general terms. In Sec. III we derive our main result, which is the emergence of unitary symplectic symmetry for two-dimensional Hamiltonians that are even under momentum reversal. In Sec. IV we classify the fermion bilinears into irreducible representations of the symmetry group, and discuss the reduction of the symmetry by the chemical potential in Sec. V. In Sec. VI we show how the finite expectation value of the fermion bilinear other than the singlet mass-term breaks the unitary symplectic group. Examples are discussed in Sec. VII.  The interacting theory that respects the unitary symplectic and spatial rotational symmetries is presented in Sec. VIII. The overlap of the orthogonal and symplectic groups, relevant to honeycomb lattice for example, is obtained in Sec. IX. Finally, we summarize our findings and outline some physical consequences in Sec. X.

\section{Orthogonal symmetry}

 Let us begin by  defining a general semimetal in two spatial dimensions with the imaginary-time low-energy action $S=\int_0 ^\beta d\tau d^2 \vec{x} L_0 $, and the non-interacting Lagrangian
\begin{equation}
L_0 = \Psi^\dagger [ 1_N \otimes (1_2 \partial_\tau + \sigma_1 F_1(\hat{\vec{p}} ) + \sigma_2 F_2(\hat{\vec{p}})) ]  \Psi, 
\end{equation}
where $\sigma_i$, $i=1,2,3$ are the standard Pauli matrices, $1_N$ is the $N\times N $ unit matrix, $\Psi = \Psi (\vec{x}, \tau)$ is a $2N$-component complex Grassmann field, and $F_i (\hat{\vec{p}})$ are Hermitian operators. We assume each function $F_i$ has a definite momentum-reversal  symmetry, i. e. to be either even or odd under the replacement $\hat{\vec{p}} \rightarrow - \hat{\vec{p}}$. $\hat{\vec{p}}= (-i\partial_1, -i \partial_2)$ is the usual momentum operator. The Lagrangian $L_0 $ represents a broad class of low-energy theories around $N$ Fermi points that involve two Hermitian anticommuting matrices, since these can always be transformed into $1_N \otimes \sigma_i$, $i=1,2$. 

$L_0$ is invariant under the internal transformation $\Psi \rightarrow (U\otimes 1_2) \Psi$, where $U\in U(N)=SU(N) \times U(1)$, and $U(1)$ represents the change of the phase of the complex fermion field. The internal symmetry of $L_0$ will shortly be seen to actually always be  larger. The internal space may consist of different valleys, spin-projections, or any other such flavor.  $L_0$ is also invariant under the space-inversion and the time-reversal, which as we will see can always be defined. It may be invariant under the $O(2)$ group of momentum rotations as well. 

Consider first the case when both operators are odd functions: $ F_i (-\hat{\vec{p}})= - F_i (\hat{\vec{p}})$, $i=1,2$. This includes the Lorentz-invariant Dirac semimetals such as graphene, when $F_i = \hat{p}_i$, as well as three-layer graphene systems with dispersion cubic in momentum. \cite{zhou}  The inversion symmetry may be represented by $I= 1_N \otimes \sigma_3$ and $\hat{p}_i \rightarrow -\hat{p}_i $, and the time-reversal by the antiunitary $T= (1_N \otimes \sigma_2 ) K$, with $K$ as complex conjugation, and $\hat{p}_i \rightarrow -\hat{p}_i $. \cite{operators}  Introducing the Majorana fermions $\chi_1$ and $\chi_2$ as $\Psi= \chi_1 - i \chi_2$ and $\Psi^\dagger = \chi_1 ^T + i \chi_2 ^T $ gives 
\begin{equation}
L_0 = \chi^T  [ 1_N \otimes (1_4 \partial_\tau +  1_2 \otimes \sigma_1 F_1(\hat{\vec{p}} )  + \sigma_2 \otimes \sigma_2 F_2(\hat{\vec{p}})) ]  \chi, 
\end{equation}
where $\chi^T = (\chi_1 ^T, \chi_2 ^T)$.  Due to the assumed oddness of the functions $F_i$, all three operators in the bracket are {\it odd under transposition}. This is a necessary condition in order for them to contribute to $L_0$, and it derives from the Grassmann nature of the Majorana variables. Since both matrices $1_2 \otimes \sigma_1$ and $\sigma_2 \otimes \sigma_2$ are symmetric and real, they are reducible to two blocks of symmetric Pauli matrices. The orthogonal transformation that accomplishes this is  
\begin{equation} 
\chi \rightarrow \exp[ i (\pi/4) (1_N \otimes \sigma_2 \otimes \sigma_1) ] \chi, 
\end{equation}
 which transforms $L_0$ into 
 \begin{equation}
L_0 = \chi^T  [ 1_{2N} \otimes ( 1_2 \partial_\tau +  \sigma_1 F_1(\hat{\vec{p}} ) +  \sigma_3 F_2(\hat{\vec{p}})) ]  \chi. 
\end{equation}
It now becomes evident that $L_0$ is invariant under a larger set of internal transformations $\chi \rightarrow   (O\otimes 1_2) \chi$, 
with $O \in O(2N)\supset U(N)$. \cite{HerbutMandal}  It should also be clear that the same enlarged orthogonal symmetry would emerge for any odd  $F_1$ and $F_2$, irrespective of their precise forms. 

Consider next the case of mixed symmetry under momentum reversal, when $ F_1 (\hat{\vec{p}})= - F_1 (-\hat{\vec{p}})$ is odd, but $F_2 (\hat{\vec{p}})= F_2 (-\hat{\vec{p}})$ is even. This, for example, is the case of two just-merged Fermi points, when $F_1 = \hat{p}_1$ and $F_2 = \hat{p}_2 ^2$. \cite{Montambaux, Dora1} The space-inversion may now be defined as $I=1_N \otimes \sigma_2$ and the time-reversal as $T= (1_N \otimes \sigma_3) K$, both accompanied with $\hat{p}_i \rightarrow -\hat{p}_i$. In terms of Majorana fermions one then finds 
\begin{equation}
L_0 = \chi^T  [ 1_{2N}\otimes (1_2 \partial_\tau + \sigma_1 F_1(\hat{\vec{p}} ) + \sigma_2 F_2(\hat{\vec{p}})) ]  \chi, 
\end{equation}
so that the symmetry group still remains $O(2N)$. The reader would obtain the same result for the reverse choice of even $F_1$ and odd $F_2$.

\section{Symplectic symmetry} 

 A different symmetry arises, however, when {\it both} operators are even functions of the momentum: $ F_i (-\hat{\vec{p}})= F_i (\hat{\vec{p}})$, $i=1,2$. An important example is provided by the low-energy theory of the Bernal-stacked bilayer graphene, when one can take $F_1= \hat{p}_1 ^2 - \hat{p}_2 ^2$ and $F_2 = 2 \hat{p}_1 \hat{p}_2$, or by hopping Hamiltonian on checkerboard and Kagome lattices. \cite{Sun}  The two discrete symmetries are now 
$I= 1_{2N}$ and $T= (1_N \otimes \sigma_1) K$, together with $\hat{p}_i \rightarrow -\hat{p}_i$.  In Majorana representation the Lagrangian becomes  
\begin{equation}
L_0= \chi^T  [ 1_N \otimes (1_4 \partial_\tau + \beta_1 F_1(\hat{\vec{p}} ) + \beta_2 F_2(\hat{\vec{p}})) ]  \chi. 
\end{equation}
The matrices $\beta_1 = \sigma_2 \otimes \sigma_1$ and $\beta_2 = 1_2 \otimes \sigma_2$ are now both {\it antisymmetric and imaginary}. This ensures that the operator in the bracket still comes out odd under transposition, as necessary. The same feature, however, makes the above representation of the two anticommuting $\beta_1$ and $\beta_2$ {\it irreducible}, since maximally one out of three anticommuting $2\times2$  matrices can be chosen to be antisymmetric.\cite{Pauli, 2012} 

The matrices $\beta_1$ and $\beta_2 $ cannot be block-reduced, but they can be complemented with three other symmetric and real $4 \times 4$ matrices $\alpha_1 = 1_2 \otimes \sigma_3$, $ \alpha_2 = \sigma_1 \otimes \sigma_1$, and $\alpha_3= \sigma_3 \otimes \sigma_1 $, so that $\alpha_i$, $i=1,2,3$ and $ \beta_j $, $j=1,2$ {\it all} mutually anticommute. That way they provide a $4\times 4$ Hermitian representation of the five-dimensional Clifford algebra $\{\gamma_i, \gamma_j \} = 2\delta_{ij}$. The three pairs of $\alpha_i$ matrices can be used to define Hermitian matrices  $X_i$, $i=1,2,3$:  
\begin{equation}
  X_i = \frac{i}{2} \epsilon_{ijk} \alpha_j \alpha_k, 
\end{equation}
 which  {\it commute} with both $\beta_1$ and $\beta_2$:
\begin{equation}
[X_i, \beta_j] = 0.
\end{equation}

Since $X_i$ are also antisymmetric and imaginary by construction, the real representation of the symmetry group of $L_0$ in Eq. (6) is now generated by the set of $4N\times 4N $ (antisymmetric, imaginary) generators $\{ A_N \otimes 1_4, S_N \otimes X_i \}$, with $S_N$ ($A_N$) as linearly independent,  real, symmetric (imaginary, antisymmetric) $N\times N$ matrices. So generated symmetry group is therefore still represented by $4N\times 4N $ real orthogonal matrices, as required by the invariance of the first term $\sim \chi^T 1_{4N}\partial_\tau \chi $ in Eq. (6). The number of the generators is, however, 
\begin{equation}
\frac{N^2 -N}{2} + 3 \frac{N^2 + N}{2}= N (2N+1), 
\end{equation}
and clearly different than the number of the generators of $O(2N)$, which is only $N (2N-1) $. We show next that the generated symmetry group is actually the {\it unitary symplectic group}, $USp(2N)$, in disguise, as the dimension of its Lie algebra in Eq. (9) already suggests. \cite{Zee}    

For that purpose, let us exhibit the matrices $X_i$: $X_1= \sigma_2 \otimes 1_2$, $X_2 = \sigma_3 \otimes \sigma_2 $, and $X_3= - \sigma_1 \otimes \sigma_2 $. The presence of $\sigma_2$ in the last factor is what makes the generators $S_N \otimes X_{2,3}$ antisymmetric, and consequently, the representation of the group real. To simply identify the symmetry group it is useful, however, to rotate that $\sigma_2 $ in the last factor into $\sigma_3$, and consider a  complex representation of the same matrix dimension instead. This transformation readily leads to a block-diagonal representation of all the generators, with the generators becoming an orthogonal sum of two $2N\times 2N$ blocks $G_a$, with $G_a \in \{ A_N \otimes 1_2, s_i S_N\otimes \sigma_i \}$, (no summation assumed) $i=1,2,3$. The signs are given as $-s_1 = s_2 = s_3 =1 $ for the upper, and $s_1 = s_2 = -s_3 = 1$ for the lower block.     

Each $2N\times 2N$ block $G_a $ therefore evidently satisfies the condition:
\begin{equation}
(1_N \otimes \sigma_2) G_a ^* (1_N \otimes \sigma_2) = -G_a.
\end{equation}
The unitary matrix $U= \exp(i \theta _a G_a)$ consequently obeys,   
\begin{equation}
(1_N \otimes \sigma_2) U (1_N \otimes \sigma_2) = U^* . 
\end{equation}
The last condition, however, is equivalent to  
\begin{equation}
U^T (1_N\otimes \tau) U = 1_N \otimes \tau, 
\end{equation}
where $\tau = i\sigma_2$. This last equation is the definition of the symplectic group of $2N\times 2N$ unitary matrices $USp(2N)$.\cite{Zee}

We conclude that the original $4N\times 4N $ matrix representation of the symmetry group of Eq. (6) is orthogonal and real, and block-reducible to the sum of two complex conjugate irreps of $USp(2N)$. 

Note that even if the Lagrangian in Eq. (1) violated  the time-reversal symmetry by containing an additional term $\Psi^\dagger [1_N \otimes \sigma_3] F_3(\hat{\vec{p}}) \Psi$ with an even function $F_3$ \cite{Dora2}, the internal symmetry would remain $USp(2N)$. In the Majorana representation this term would translate into $\chi^T [1_N \otimes i\beta_1 \beta_2  F_3(\hat{\vec{p}}) ] \chi$, and since $[X_i , i \beta_1 \beta_2]=0$, the set of symmetry generators remains unchanged.

\section{Fermion bilinears}

 It is now straightforward to classify all Majorana bilinears $\chi^T Y \chi$, with the matrix $Y$ satisfying $Y^T  = - Y$, as irreps of the $USp(2N)$.
 
First, if the matrix $Y$ belongs to the Lie algebra of $USp(2N)$, the corresponding bilinear is in the ``adjoint" irrep $N(2N+1)$. Second, consider  $Y=M$, with $M$ as a $4N\times 4N $ mass-matrix that anticommutes with the Majorana Hamiltonian: $\{ M, 1_N \otimes \beta_i \}=0$, $i=1,2$. The requirement of antisymmetry of $M$ implies that  
\begin{equation}
M \in \{ S_N \otimes i\beta_1 \beta_2, A_N \otimes \alpha_i\}.
\end{equation}
Keeping in mind that $\alpha_1 \alpha_2 \alpha_3 \beta_1 \beta_2 = 1_4$ and by computing the commutators of the mass-matrices one finds that these $N(2N-1)$ matrices do not close a Lie algebra by themselves, but together with the $N(2N+1)$ generators of the $USp(2N)$ form the Lie algebra $U(2N)$. One of the matrices, namely $M_0 = 1_N \otimes i\beta_1 \beta_2 $, commutes with all the generators, and is therefore a scalar under $USp(2N)$. Since it also commutes with all of the remaining mass-matrices, Schur's lemma implies that the rest is a block-reducible representation of the $USp(2N)$. 

The transformation $ M \rightarrow U M U^\dagger$ with  $U= \exp[ i (\pi/4) (1_N \otimes (1_2 + \sigma_2) \otimes \sigma_1)] $ reduces the masses $M \neq M_0$ to a block-diagonal form, with the upper block as $\{ S_N \otimes 1_2, A_N \otimes \sigma_1, A_N \otimes \sigma_2, A_N \otimes \sigma_3 \}$, and the lower block as  $\{- S_N \otimes 1_2, A_N \otimes \sigma_1, -A_N \otimes \sigma_2, A_N \otimes \sigma_3\}$. Each $2N\times 2N $ block therefore obeys the same condition in Eq. (11).

The two blocks are two complex-conjugate irreps of $USp(2N)$ of dimensions $N(2N-1) -1$. This is in accord with the analogous conclusion already drawn about the blocks in the generators of the $USp(2N)$. Altogether therefore, the block-reducible $4N \times 4N$ mass-matrices $M\neq M_0$ in Eq. (13)  transform under the block-reducible $4N \times 4N$ real representation of $USp(2N)$ as $M\rightarrow U M U^\dagger$. 

Finally, there are two further $N(2N-1)$-dimensional representations given by $Y= (1_N \otimes \beta_i)M$, $i=1,2$, to which the same considerations apply. The corresponding fermion bilinears may be called ``nematics" \cite{Sun}, since their finite expectation value would break the spatial rotational symmetry $O(2)$. The generators of $USp(2N)$, the masses, and the nematics exhaust the set of  possible Majorana bilinears.

\section{Finite chemical potential} 

We now add a term $\chi^T G \chi$ with $G$ as one of the $USp(2N)$ generators to the Lagrangian in Eq. (6). An important example is $G_0=1_N\otimes X_1$, which represents the particle number operator, so that the corresponding bilinear couples to the chemical potential. $G_0$  commutes only with the generators $\{ A_N \otimes 1_4, S_N \otimes X_1 \}$. Diagonalizing  $X_1$ one readily finds that this Lie sub-algebra is an orthogonal sum of four blocks of $\{A_N, \pm S_N \}$, each of which generates the unitary group $U(N)$. A finite chemical potential therefore restricts $USp(2N)$ to $U(N)= SU(N)\times U(1)$, with $U(1)$ generated by $G_0$. This is nothing but the ``original" $U(N)$ discussed below Eq. (1).  

Under the restriction of $USp(2N)$ to $U(N)= SU(N)\times U(1)$ the irrep of mass-matrices $N(2N-1)-1$ of $USp(2N)$ breaks into the irreps of $SU(N)$ as  
\begin{equation}
N(2N-1)-1 = (N^2-1) + \frac{N(N-1)}{2} + \overline{\frac{N(N-1)}{2} },
\end{equation}
where the irrep ``$N^2 -1$" consists of $\{ S_N \otimes i\beta_1 \beta_2, A_N\otimes \alpha_1 \}$  ($S_N \neq 1_N$) and transforms as the adjoint of $SU(N)$ and as a singlet under $U(1)$. The irreps ``$N(N-1)/2$" and ``$\overline{N(N-1)/2}$" consist of $\{ A_N\otimes \alpha_2 \}$ and  
$\{ A_N \otimes \alpha_3 \}$, respectively. Each transforms as the antisymmetric second-rank tensor under $SU(N)$, and together they form a doublet under $U(1)$. The $U(1)$ singlets represent the gapped insulators, and the doublets gapped superconductors. Their respective numbers agree with those found in Ref. \cite{Roy}.  The nematic irreps follow an analogous pattern.

\section{Symmetry breaking by a mass-term} 

Consider next a finite average $\langle \chi^T M \chi \rangle \neq 0$, with  $M\neq 1_N \otimes i\beta_1 \beta_2 $, arising either by spontaneous or explicit symmetry breaking. If we assume that $M^2 = 1_{4N}$, so that all fermion components have the same magnitude of the gap, \cite{remark2} as well as even $N$, \cite{remark3} the condition $Tr M = 0$ implies that $M$ can be transformed into block-diagonal form $\sigma_3 \otimes 1_{2N} $. The symmetry is therefore broken as 
\begin{equation}
USp(2N) \rightarrow USp(N) \times USp (N),  
\end{equation}
which entails, in particular, $N^2$ Goldstone bosons. The rotational symmetry is preserved. In the case of nematic order  
$\langle \chi^T \beta_i M \chi \rangle \neq 0$, both the internal and the rotational symmetries are broken. If $M= 1_N \otimes i\beta_1 \beta_2$ on the other hand, the internal symmetry is preserved.

\section{Examples} 

 In the case of a single quadratic-band-touching (QBT) \cite{Sun} with spinless fermions, $N=1$ and the symmetry group is $USp(2)= SU(2)$. The mass-matrix is unique and represents the quantum anomalous Hall state. There are also two nematics. All three states are singlets under $USp(2)$.    

$N=2$ can be understood as the single QBT Hamiltonian for spin-1/2 particles, \cite{Sun} or as the spin-polarized Bernal-stacked bilayer 
graphene with two QBTs.  The group $USp(4)$ is ten-dimensional and equivalent to the spinor $SO(5)$, i.e. $Spin(5)$.  
The six possible masses fall into the singlet or the five-dimensional vector representations of $Spin(5)$, with two superconducting and three insulating masses in the latter.  For the spin-1/2 single-QBT Hamiltonian the latter five represent two components of the s-wave superconductor and the three quantum spin Hall masses, which form a singlet and a vector under the spin-$SU(2)$, and a doublet and a singlet under the particle-$U(1)$, respectively. \cite{Szabo}

For $N=4$, which would correspond to physical Bernal-stacked bilayer graphene with spin-1/2 electrons, $USp(8)$ is 36-dimensional, and there are $1+27$ different mass-terms. Among the 27 symmetry-breaking mass-order-parameters, 15 represent insulators, and the remaining 12 gapped superconductors. \cite{Roy} There are also twice as many nematic states, which would split QBT into Dirac points. 

\section{Interactions}

 To determine the minimal interacting theory we take the QBT non-interacting Hamiltonian to be invariant under the $O(2)$ of spatial rotations generated by $1_N\otimes \sigma_3$, i. e. by $1_N \otimes i \beta_1 \beta_2 $ in Majorana representation, accompanied by a rotation of the momentum. It was shown before that in the Dirac case there are {\it two}  linearly independent $U(N)$-symmetric quartic terms. \cite{HerbutMandal} It is easy to check that the restriction of the Lorentz ($O(2+1)$) to the rotational ($O(2)$) symmetry increases this number to {\it three}. For reasons that will become clear shortly, we take these three to be: \cite{remark}
\begin{equation}
I_1 = (\Psi^\dagger 1_N \otimes \sigma_3 \Psi )^2, 
\end{equation}
 \begin{equation}
I_2 = (\Psi^\dagger H \otimes \sigma_3 \Psi )^2 , 
\end{equation}
\begin{equation}
I_3 = (\Psi ^T A_N \otimes \sigma_1 \Psi ) (\Psi ^\dagger A_N \otimes \sigma_1 \Psi^* ), 
\end{equation}
with the summation over all $N^2 -1$ linearly independent Hermitian generators $H$ of $SU(N)$ in Eq. (17), and over all imaginary, antisymmetric,  linearly independent $N\times N$ matrices $A_N$ in Eq. (18) assumed. The squared bilinears appearing in Eqs. (16) - (18)  belong to the singlet-singlet, adjoint-singlet, and antisymmetric tensor-singlet representations of $SU(N) \times O(2)$, respectively. In terms of Majorana fermions, 
\begin{equation}
I_1 = (\chi^T 1_N \otimes i\beta_1 \beta_2 \chi )^2, 
\end{equation}
\begin{equation}
I_2 = (\chi^T A_N \otimes \alpha_1 \chi )^2 +  (\chi^T S_N \otimes i \beta_1 \beta_2 \chi )^2,
\end{equation}
\begin{equation}
I_3 = (\chi^T A_N \otimes \alpha_2 \chi )^2 +  (\chi^T A_N \otimes \alpha_3 \chi )^2,
\end{equation}
with the same summation convention as in Eqs. (16)-(18) extended to the $N\times N$ symmetric matrices $S_N$ as well. 
The quartic term $I_1$  is clearly invariant under $USp(2N)$, whereas $I_2$ and $I_3$ individually are not. However, their sum is:   
\begin{equation}
I_2 + I_3 =  (\chi^T M \chi )^2,
\end{equation}
with the summation over all masses $M\neq 1_N \otimes i \beta_1 \beta_2 $ displayed in Eq. (13). $I_2 + I_3$ is therefore the second independent quartic term invariant under $USp(2N)$. The general $U(N) \times O(2)$-symmetric interacting Lagrangian is therefore       
\begin{equation}
L = L_0+ g_1 I_1 + g_2 I_2 + g_3 I_3, 
\end{equation}
where $g_i$ represent the interaction coupling constants. When $g_2 = g_3$, the interacting theory given by $L$ becomes $USp(2N)\times O(2)$-symmetric. 

By power counting, all three coupling constants in Eq. (23) are marginal, but the loop corrections make them marginally irrelevant or marginally relevant. \cite{Szabo, Vafek} If they are marginally relevant, the internal symmetry may either be preserved or become broken as in Eq. (15), in either a massive or a nematic ground state. The detailed analysis will be presented in a separate paper.

\section{Asymmetric Hamiltonian}

If any of the functions $F_i (\hat{\vec{p}} )$ had no definite symmetry under momentum reversal, neither the space-inversion  nor the time-reversal operators could be defined, and both symmetries must be broken. An immediate implication is that if a space-inversion-symmetric or a time-reversal-symmetric lattice Hamiltonian exhibits a single QBT in the Brillouin zone, the entire Hamiltonian, and not just its leading term, must be an even function of the momentum. In that case $USp(2N)$ becomes an {\it exact} symmetry. An example is the tight-binding Hamiltonian on the checkerboard lattice of Ref. \cite{Sun}. 

To find both momentum-reversal even and odd terms in the Hamiltonian there have to exist at least two Fermi points, like in graphene. The non-interacting Lagrangian, for even $N$ and in terms of the complex fermions, can then be written as
\begin{widetext}
\begin{equation}
L_0 = \Psi^\dagger [ 1_N \otimes (1_2 \partial_\tau + \sigma_1 F_1 ^{od} (\hat{\vec{p}} ) + \sigma_2 F_2 ^{od} (\hat{\vec{p}})) 
+ 1_{N/2} \otimes \sigma_3 \otimes ( \sigma_1 F_1 ^{ev} (\hat{\vec{p}} ) + \sigma_2 F_2 ^{ev} (\hat{\vec{p}})) ] \Psi, 
\end{equation}
\end{widetext}
where $F_i ^{ev}$ ($F_i ^{od}$ ) are even (odd) functions. Note how the matrix $1_{N/2} \otimes \sigma_3$ replaces  $1_N$ in the second term,  which reflects the opposite ``chiralities" of the two Fermi points. This is precisely what enables one to define the time-reversal operator as $T= (1_{N/2} \otimes \sigma_2 \otimes \sigma_2 )K$, and the inversion as $I= 1_{N/2} \otimes \sigma_2 \otimes \sigma_3$, for example. \cite{choice, HerbutJuricicRoy}  
 With both even and odd terms present, however, the original  Lie symmetry group $U(N)$ of Eq. (1) is reduced to $U(N/2)\otimes U(N/2)$. For spin-1/2 electrons ($N=4$), for example, this group consists of the rotations of particle spin and change of the phase for each Fermi point separately, but excludes the ``rotations" between the Fermi points. 

We show next that, surprisingly, the Lagrangian in Eq. (24) nevertheless acquires a {\it new} $U(N)$ symmetry. The first term in Eq. (24) already has the form of Eq. (1), and therefore exhibits $O(2N)$ symmetry in the Majorana representation. The second term can also be easily transformed into the form as in Eq. (1), and therefore alone will show $USp(2N)$ symmetry in Majorana representation. The symmetry of the sum of the two terms in Eq. (24) is therefore the {\it overlap} of these two groups. We show next that this is yet another $U(N)$, different from the ``original" $U(N)$. 

The Lagrangian in the last equation in Majorana representation becomes 
\begin{widetext}
\begin{equation}
L_0 = \chi^T  [ 1_N \otimes (1_4 \partial_\tau + i\alpha_1 \beta_2 F_1 ^{od} (\hat{\vec{p}} ) - i \alpha_1 \beta_1 F_2 ^{od} (\hat{\vec{p}}) ) 
+ 1_{N/2} \otimes \sigma_3 \otimes ( \beta_1 F_1 ^{ev} (\hat{\vec{p}} ) + \beta_2 F_2 ^{ev} (\hat{\vec{p}}) ) ] \chi, 
\end{equation}
\end{widetext}
where we wrote the two matrices appearing in Eq. (2) as $1_2 \otimes \sigma_1 = i \alpha_1 \beta_2$ and  $\sigma_2 \otimes \sigma_2 = -i \alpha_1 \beta_1$. 
A symmetry generator in the desired overlap commutes therefore both with the two matrices in the first (odd) term,  as well as with the two matrices in the second (even) term. It therefore commutes with their respective two-products, which conveniently are the same for both pairs:
\begin{widetext} 
\begin{equation}
(i 1_N \otimes (\alpha_1 \beta_2)) (-i 1_N \otimes (\alpha_1 \beta_1) ) = (1_{N/2} \otimes \sigma_3 \otimes \beta_1 ) (1_{N/2} \otimes \sigma_3 \otimes \beta_2 ) = i 1_N \otimes \sigma_2 \otimes \sigma_3. 
\end{equation}
\end{widetext}
The final matrix on the right-hand side can be transformed into $1_{2N}\otimes \sigma_3$, and in that particular basis the generators of both symmetries $O(2N)$ of the first and $USp(2N)$ of the second term become orthogonal sums of two  $2N \times 2N$ unitary blocks. 

Each such block therefore simultaneously satisfies the orthogonal and the symplectic conditions. In a general basis the orthogonal condition on the generator is 
\begin{equation}
U_1 G^*   U_1 ^{-1} = -G,
\end{equation}
with the unitary $U_1$ satisfying $U_1 U_1 ^*= 1$. The condition can be understood as the generator being odd under an antiunitary operation $A_1=U_1 K$, such that $A_1 ^2 = +1$. The general symplectic condition on the generator, on the other hand, is 
\begin{equation}
U_2 G^* U_2  ^{-1} = -G,
\end{equation}
 with $U_2 U_2 ^* = -1$. This condition can similarly be understood as the generator being odd under an antiunitary operation $A_2 = U_2 K$, with $A_2 ^2 = -1$.  To find the overlap between the two conditions we may first choose the basis where $U_1 = 1_{2N}$.\cite{orthogonal} In this basis $G^* = -G$, and the symplectic condition becomes simply 
 \begin{equation}
 [G, U_2] =0. 
 \end{equation}
 Now we may change the basis into one in which $U_2 = 1_N \otimes \sigma_2$, as in Eq. (10). \cite{symplectic} In this basis, among all solutions of the symplectic condition in Eq. (28),  $\{ A_N \otimes 1_2, S_N \otimes \sigma_i \}$,  the only ones that satisfy Eq. (29) are $\{ A_N \otimes 1_2, S_N \otimes \sigma_2 \}$. Since $\sigma_2$ can be rotated into $\sigma_3$ we see that each $2N \times 2N$ block is an orthogonal sum of two $N\times N$ blocks of matrices $\{ A_N, \pm S_N\}$, which, for either sign, are nothing but the generators of $U(N)$. The final result is that the original $4N\times 4N$ representation of the symmetry group of the Lagrangian in Eq. (25) reduces to two pairs of complex conjugate representations of $U(N)$. In other words, $O(2N) \bigcap USp(2N) = U(N)$, as announced.  
 
 \section{Summary and discussion} 
 
 We showed that the symmetry of the QBT Hamiltonian in two dimensions is the unitary symplectic group, and we used it to classify fermion bilinears, discuss possible symmetry breaking patterns, and construct the minimal interacting theory that would respect it. The symmetric interacting theory may be understood as an equivalent of the relativistic Gross-Neveu field theory for the non-relativistic QBT, and in contrast to the relativistic case, it generally 
 features two independent interaction terms. On lattices such as honeycomb or Bernal-stacked graphene bilayer the symmetry to all orders in the expansion of the energy dispersion in powers of local momentum is given by the overlap of the orthogonal and unitary symplectic symmetries,  which we demonstrated yields another unitary symmetry. This is the first time to our knowledge that a symplectic group emerges as a group of symmetry of a quantum Hamiltonian.

 Just as in the case of $O(2N)$ symmetry of the Dirac theory \cite{HerbutMandal}, maybe the most interesting feature of the $USp(2N)$ symmetry is that it contains generators that anticommute with the particle-number operator, and rotate the insulating into the superconducting fermion bilinears, some of which belong to the same irrep. If the symmetry happens to be weakly broken in favor of an insulator at zero chemical potential, for example, its increase  will ``flop" the order parameter into a superconducting direction.   
  
 The Bernal-stacked bilayer with trigonal warping suppressed has a QBT as the leading term in the momentum expansion, followed by higher-order terms starting from the cubic. The $USp(8)$ therefore emerges here only in the low-energy limit, even with the electron-electron interactions neglected. A better example of the $USp(2N)$ symmetry is the tight-binding Hamiltonian on the checkerboard lattice of ref. \cite{Sun}, which has the energy dispersion as an even function of the momentum. This feature, as we have seen, implies the unitary symplectic symmetry to all orders in the momentum expansion. In the presence of only short-range interactions, the quartic terms that are marginal by power counting can in general be represented as linear combinations of those in Eq. (23). \cite{Fierz} In a rich-enough lattice model it therefore should be possible to fine tune the interaction to a $USp(2N)$-symmetric form. It is also conceivable that the combination $g_2 - g_3$, when small, is an irrelevant perturbation, and that at least for some initial conditions the unitary symplectic symmetry emerges from  the renormalization group flow towards the infrared.  In this scenario, if the $USp(2N)$ turns out to be spontaneously broken down to $USp(N) \times USp(N)$, some of the ensuing $N^2$ Goldstone bosons would acquire a small gap. A weak $g_2 - g_3$ perturbation would reduce the original $USp(2N)$ to $U(N)$, and therefore the number of gapped Goldstone bosons from Eq. (14) would be $N^2 - (N^2 -2)=2$  or $ N^2 - ( N^2 -N-1) = N+1 $, depending on whether the broken symmetry state is an insulator or a superconductor. Detection of these light collective modes in the ordered state would be a sign of the spontaneously broken infrared-emergent $USp(2N)$.
 
It is also interesting to ponder when more generally one may expect the appearance of the unitary symplectic symmetry in the low-energy single-particle Hamiltonian. Our construction identifies the crucial ingredient to be the existence of the three non-trivial matrices $X_i$ that commute with the matrices $\beta_{1,2}$ which appear in the Hamiltonian in the Majorana representation. More formally, the matrices $i\beta_{1,2}$ close the Clifford algebra $Cl(0,2)$, \cite{Clifford} and define its real, unique, $4 \times 4$ irrep. The important point is that this irrep is {\it quaternionic}: it has four real Casimir operators, $\{1_4, i X_i  \}$, which close the quaternionic algebra. \cite{Pauli, 2012} Another example of a quaternionic real representation is provided by $Cl(4,0)$, which, following the logic of Sec. II, appears in the Dirac Hamiltonian in four spatial dimensions. Such a Dirac Hamiltonian, in spite of being odd under momentum reversal, therefore also admits the unitary symplectic symmetry. The connection between the algebraic  nature of the real irreps of Clifford algebras that appear in the Majorana representations of the semimetals' Hamiltonians and their hidden symmetries will be further elaborated in a separate publication.  

Symplectic group has also recently appeared in Ref. \cite{Masaoka}, as a symmetry of the auxiliary theory which generates equal-time correlators in the 
non-interacting QBT system in two dimensions. Its origin there is rather different from that of our unitary symplectic group, which can be the symmetry of the interacting theory as well, as well as of the massive QBT Hamiltonian, and as we argued above, even of the Dirac theory in four spatial dimensions.

\section{Acknowledgements}

 The authors acknowledge useful discussions with Fakher Assaad, SangEun Han, Rintaro Masaoka, and Bitan Roy. This work has been supported by the NSERC of Canada.


\begin{thebibliography}{99}
\bibitem{irrep} Here we will call an irrep by its dimension.
\bibitem{operators} The choice of operators that represent the inversion and the time-reversal is obviously not unique.
\bibitem{remark3} $N$ does not have to be even since the Nielsen-Ninomiya theorem does not apply to QBT. We assume it nevertheless to include the cases of spin-1/2 and spinless graphene bilayer.
\bibitem{remark} The choice of the three linearly independent quartic terms is not unique, but cannot be made completely freely either.
\bibitem{choice} This specific choice is made so that $T$ and $I$ commute, and that $T^2=1$, but it is still not unique.
\bibitem{orthogonal} A unitary $U_1$ such that $U_1 U_1 ^* = 1$ can always be transformed into unit operator.
\bibitem{symplectic} Such a basis also can always be found.
\bibitem{Clifford} Clifford algebra $Cl(p,q)$ is defined as $p+q$ mutually anticommuting matrices, $p$ ($q$) of which square to plus  (minus) unity.
\end{thebibliography}
\end{document}